\documentclass{aa}
\usepackage{graphicx}
\usepackage{times}
\begin{document}

\thesaurus{08     
              (09.16.1;  
               13.09.6;  
               08.16.4;  
               08.03.4;  
               02.18.7)} 
\title{The dust content of planetary nebulae: a reappraisal}


\author{Gra{\.z}yna Stasi{\'n}ska\inst{1}, Ryszard Szczerba\inst{2}}

\offprints{Gra{\.z}yna Stasi{\'n}ska (grazyna@obspm.fr)}

\institute{DAEC, Observatoire de Paris--Meudon, 92195 Meudon Cedex, 
France
           {[email: grazyna@obspm.fr]}
\and
          N. Copernicus Astronomical Center, Rabia\'{n}ska 8, 
          87--100 Toru\'{n}, Poland {[email: szczerba@ncac.torun.pl]}
          }

\date{Received date; accepted date}

\titlerunning{The dust content of planetary nebulae}
\authorrunning{G. Stasi{\'n}ska \& R. Szczerba}

\maketitle

\begin{abstract}

We have performed a statistical analysis using broad band IRAS data on
about 500 planetary nebulae with  the aim of characterizing their dust
content. Our approach is  different from previous  studies in that  it
uses an extensive  grid of photoionization models  to test the methods
for deriving the dust temperature,  the dust-to-gas mass ratio and the
average grain size.   In addition, we  use only  distance  independent
diagrams.
 
With our models,  we show the  effect of contamination by atomic lines
in the broad band  IRAS fluxes during  planetary nebula evolution.  We
find that planetary nebulae with    very different dust-to-gas    mass
ratios exist, so that  the dust content is  a primordial parameter for
the  interpretation of far  infrared  data of  planetary nebulae.   In
contrast with previous studies, we find  no evidence for a decrease in
the dust-to-gas  mass ratio as the planetary  nebulae evolve.  We also
show   that  the decrease in     grain  size advocated   by  Natta  \&
Panagia~(\cite{NattaPanagia}) and   Lenzuni et al.~(\cite{Lenzuni}) is
an artefact of their method of analysis.  Our results suggest that the
timescale  for  destruction  of dust  grains  in planetary  nebulae is
larger than their lifetime.
\footnote{Table 1  is only accessible in  electronic form  at  the CDS via
anonymous   ftp   to   cdsarc.u-strasbg.fr   (130.79.128.5)   or   via
http://cdsweb.u-strasbg.fr/Abstract.html}


\end{abstract}


\section{Introduction}

Far   infrared   measurements of  planetary    nebulae   allow one  to
investigate  their dust content.  From broad band  photometry, one can
derive the characteristic  temperature of  the  emitting dust and  the
total amount of energy that it radiates.

For  example,  Natta  \&   Panagia~(\cite{NattaPanagia}),  using   the
aircraft observations by  Moseley~(\cite{Moseley}),  analyzed a sample
of 10 planetary nebulae, and found  evidence that the dust-to-gas mass
ratio,    $m_{\mathrm  d}/m_{\mathrm g}$, and   the    grain size were
decreasing systematically  with   nebular  radius,  indicating a  time
variation of the dust properties in  planetary nebulae. Then, Pottasch
et al.~(\cite{Pottasch84a}) used IRAS broad band photometry to analyze
a  sample   of 46  planetary  nebulae.   They   also found $m_{\mathrm
d}/m_{\mathrm  g}$ to decrease with  radius, being  as  large as a few
10$^{-2}$ for the smallest nebulae and two order of magnitudes smaller
for the largest ones. Later,  Lenzuni et al.~(\cite{Lenzuni}) repeated
the   analysis of Natta  \&   Panagia~(\cite{NattaPanagia}) on a  much
larger sample:  they analyzed  233  planetary nebulae  which  had been
observed by IRAS  in at least the 25\,$\mu$m  and 60\,$\mu$m bands and
appeared in the Daub~(\cite{Daub})  catalog of distances. Essentially,
they confirmed the finding of Natta \& Panagia~(\cite{NattaPanagia}).

The  problem of the infrared  luminosity of planetary nebulae has been
addressed by  several authors: Pottasch  et  al.~(\cite{Pottasch84a}),
Iyengar~(\cite{Iyengar}), Zijlstra et al.~(\cite{Zijlstra}), Ratag  et
al.~(\cite{Ratag90}),   Amnuel~(\cite{Amnuel}).       Pottasch      et
al.~(\cite{Pottasch84a}) found that very  high infrared excesses (IRE,
i.e. the infrared luminosity in terms of what  can be accounted for by
absorption of the Ly$\alpha$   radiation produced in the nebula)  were
usually  found  in the  small,  high  dust  temperature  nebulae. They
concluded that, in these cases, dust is  probably mostly heated by the
radiation of  the  central star on  the  longwave side  of Ly$\alpha$,
since   the  nebulae are young.   Ratag  et  al.~(\cite{Ratag90}) also
considered the  possibility that  high IRE  could be  due to high dust
contents  or   to  heating  by   the   interstellar  radiation  field.
Amnuel~(\cite{Amnuel}) favored the latter explanation.

All  the  studies above  based their  conclusions  either on a simple,
empirical  interpretation of  the  data, or,   in  the best  cases, on
semi-analytical   models  of  planetary nebulae  that  involve drastic
simplifications.  Also,  excepting the studies specifically devoted to
planetary nebulae in the Galactic bulge (Ratag et al.~\cite{Ratag90}),
they rely heavily on the adopted distances  to the nebulae. Therefore,
they  must  be  considered with  caution.   A safer approach,   in our
opinion,   for  studies of  statistical  nature,   is to interpret the
observed data with the help of grids of numerical models that are able
to compute the observables and use distance independent diagrams. Such
an  approach  for  the  interpretation  of   the infrared emission  of
planetary  nebulae   has been    attempted by  Volk~(\cite{Volk})  and
Vil'koviskii  \& Efimov~(\cite{Vil'koviskii}).   Unfortunately,   both
these studies were rather exploratory,  and did not address explicitly
the questions mentioned above.

Fifteen   years  after  the publication  of  the  results  of the IRAS
mission,  there is still work  to be done   on their interpretation as
regards  the statistical dust properties of   planetary nebulae. It is
the purpose of  the present paper  to contribute to this, by analyzing
the broad band IRAS fluxes of planetary nebulae in the light of series
of dusty photoionization  models for evolving planetary  nebulae using
distance independent diagrams. In  Section  2, we briefly present  the
observational material. In Section  3, we describe the photoionization
models.  In Section 4, we analyze the contamination of IRAS broad band
fluxes    by  line  emission.  In Section    5,   we discuss the  dust
temperature. Section  6  is devoted to   the  dust-to-gas mass  ratio,
Section  7 to the infrared  excess and Section  8 to the grain size. A
general discussion of the main findings is given in Section 9.


\section{The observational data base}

We have  considered  the planetary   nebulae from the   Strasbourg-ESO
Catalogue of Galactic  Planetary Nebulae (Acker et al.~\cite{Acker92})
that appear in the IRAS Point Source Catalog~(\cite{IRASPSC}) and have
flux   quality   $Q$\,=\,3  in   at  least    the  25  and  60\,$\mu$m
bands. According to  the IRAS Point Source Catalog,  the error in  the
fluxes with  $Q$\,=\,3 is  about 10\,\%. There  are  548 such objects.
The inclusion of planetary nebulae with $Q$\,=\,2 adds only 86 objects
with lower quality  data, and does not  change the conclusions of this
paper.

Since individually   determined  distances  in planetary nebulae   are
scarce and are sometimes contradictory  for the same object when using
different  methods (e.g. Sabbadin~\cite{Sabbadin}),  we consider  only
distance-independent diagrams. We use the reddening corrected H$\beta$
fluxes (or, equivalently, the   radio fluxes at 6\,cm   converted into
H$\beta$) and  angular diameters from the  same sources as  Tylenda \&
Stasi\'{n}ska~(\cite{Tylenda}).  We discard data  that are  only upper
limits. We  end up  with 477  nebulae  for which  we can   compute the
average surface brightness, $S$(H$\beta$), and which have the required
IRAS data   (i.e. 25 and  60\,$\mu$m  fluxes with   $Q$\,=\,3).  Table
1\footnote{Table 1  is accessible in  electronic form  at  the CDS via
anonymous   ftp   to   cdsarc.u-strasbg.fr   (130.79.128.5)   or   via
http://cdsweb.u-strasbg.fr/Abstract.html}  lists, in  columns 3 and 4,
the   adopted values of the reddening    corrected H$\beta$ fluxes and
angular radii for all the nebulae having 25 and 60\,$\mu$m fluxes with
$Q$\,=\,3  (columns   1    and 2   give   the  planetary  nebula  PN~G
identifications and usual names, respectively).  As is known, and seen
from  our models below, $S$(H$\beta$)  is a rather robust estimator of
the planetary  nebula age. It  is safer  than the  estimated planetary
nebula radius,  which was  used in former  studies on  the dust-to-gas
mass ratio  in planetary nebulae,  among other reasons,  because it is
distance independent.

Finally, to discuss the dust-to-gas mass ratio, we need an estimate of
the gas  density. We take   the  electron densities derived from   the
[S\,II] ratios as compiled by G\'{o}rny (private communication), again
discarding upper or lower limits. These densities are listed in Column
5 of Table 1.  The main references are the compilation of Stanghellini
\& Kaler~(\cite{Stanghellini}), and the observations reported by Acker
et al.~(\cite{Acker89}), Acker  et al.~(\cite{Acker91}) and Kingsburgh
\& Barlow~(\cite{Kingsburgh}).   Thus, we end up  with a sample of 230
objects for which  we can  discuss the  time evolution of  $m_{\mathrm
d}/m_{\mathrm  g}$.  This sample  is similar in    size to the  one of
Lenzuni et al.~(\cite{Lenzuni}),  but differs slightly from it because
of our more restrictive conditions on  the IRAS fluxes and more recent
compilation  of   H$\beta$     fluxes and  angular    diameters   than
Daub~(\cite{Daub}).


\section{The photoionization models}

The modelling   follows the   same   policy as  in   Stasi{\'n}ska  et
al.~(\cite{Stasinska98}).  We construct  sequences of  photoionization
models that represent the evolution of a  nebula around a central star
that has left  the  AGB and is   progressing towards the white   dwarf
stage. The nebula is modelled as a uniformly expanding spherical shell
of given    total   mass  (with a  relative    thickness  $(R_{\mathrm
{out}}-R_{\mathrm {in}})/R_{\mathrm {out}}$\,=\,0.3, where $R_{\mathrm
{in}}$ and $R_{\mathrm  {out}}$ are the inner  and outer radius of the
nebula,  respectively) in which the  density  is supposed uniform.  In
such a   simple representation,  the parameters  defining  a planetary
nebula   are  the  central  star mass,   $M_{*}$,    the nebular mass,
$M_{\mathrm {neb}}$, its  expansion velocity, $v_{\mathrm {exp}}$, and
its chemical composition (also assumed uniform).  True nebulae are, of
course, more   complicated  than  such  a  simple  description.   More
elaborate dynamical   models   have been   computed,   even  including
departure  from       spherical      symmetry  (e.g.      Frank     \&
Mellema~\cite{Frank}), which aim at  a  better representation of   the
real  evolution of planetary nebulae  (but those models do not include
dust and  only  a few sequences are  computed).  Our  approach has the
virtue  of being sufficiently  simple to  allow  one to run  series of
models    covering  the  whole  parameter    space  suggested  by  the
observations and  to  make easier the interpretation  of observational
data, in the framework of general  ideas on the evolution of planetary
nebulae.

The models are computed with the photoionization code PHOTO, using the
atomic           data       described     in        Stasi\'{n}ska   \&
Leitherer~(\cite{Stasinska96}).  The  code assumes spherical geometry,
with a central   ionizing source.  The  diffuse  radiation is  treated
assuming that all the photons are emitted outwards in a solid angle of
2$\pi$, and  the  transfer of the   resonant photons  of  hydrogen and
helium  is computed  with   the same outward  only approximation,  but
multiplying  the  photo-absorption  cross-section  by  an  appropriate
factor to   account for the  increased path  lenght due  to scattering
(Adams~\cite{Adams}).  In contrast  with  the model planetary  nebulae
presented  in  Stasi\'{n}ska  et al.~(\cite{Stasinska98}), which  were
aimed at  discussing the emission  from the  ionized  region only, the
present  models  must take  into  account  the  possible presence of a
neutral zone surrounding the ionized region,  where dust emission also
occurs. This  is done  by  extending the computation  of the radiation
transfer  outside the   ionized  zone, until   the  total nebular mass
assigned  to a given  model is   reached. The  gas temperature is  not
computed in the neutral zone, but simply set to 100K,  since it has no
incidence on the results presented here.

Dust    is   introduced    as       described  in Harrington        et
al.~(\cite{Harrington}).   In short, it is  assumed  to be composed of
spherical    grains with the   classical   size distribution ($\propto
a^{-3.5}$,   where   $a$    is   the  grain    size)     of Mathis  et
al.~(\cite{Mathis}). At each nebular radius, the equation of radiative
equilibrium  is solved,  for  18  grain sizes  logarithmically  spaced
between   0.005  and 0.25  $\mu$m. Dust  in  the  models is assumed to
interact only  with the photons,  not  with the  gas particles.  It is
heated   by the attenuated  stellar  radiation  field  and the diffuse
radiation field produced  in the  nebula.  At  each radius,  once  the
grain   temperature   is known  for   each  grain   size, the infrared
emissivity  is computed at  256 wavelenghts.   The total luminosity at
each  wavelength is   calculated    by integration over the    nebular
volume. The luminosities in the IRAS bands are obtained by convolution
with the response function of the IRAS  filters.  As a check, the same
procedure  has been  applied to  the  photoionization code written  by
Szczerba (see  e.g.  G{\c{e}}sicki et  al.~\cite{Gesicki}). Both codes
give consistent results,  and  are  able to  reproduce  the models  of
IC\,418 and NGC\,7662 by Hoare~(\cite{Hoare}).

The central stars are  assumed to evolve   according to the  H-burning
post-AGB  evolutionary   tracks   as   interpolated by   G{\'o}rny  et
al.~(\cite{Gorny})   from the tracks of Bl{\"{o}}cker~(\cite{Blocker})
and references   therein.   The  stars are   assumed   to  radiate  as
blackbodies.

Results    are  presented   with  only  one    nebular abundance   set
(He/H\,=\,0.084;    C/H\,=\,8.2\,10$^{-5}$;    N/H\,=\,2.2\,$10^{-5}$;
O/H\,=\,4.2\,$10^{-4}$;\,Ne/H\,=\,8.5\,$10^{-5}$;\,Mg/H\,=\,2.1\,$10^{-5}$;
\,Si/H\,=\,2.2\,$10^{-5}$;\,S/H\,=\,1.0\,$10^{-5}$;\,Cl/H\,=\,1.6\,$10^{-7}$;
\,Ar/H\,=\,3.0\,$10^{-6}$;\,Fe/H\,=\,2.4\,$10^{-6}$), corresponding to
half solar metallicity and depleted metals.  In the present study, the
only important effect of the gas chemical composition is to affect the
energy emitted in the lines included in the IRAS bands (see below).

IRAS   Low   Resolution    Spectroscopy of     planetary  nebulae (see
e.g.  Pottasch~\cite{Pottasch87})  has shown  that  planetary  nebulae
contain several kinds of dust grains.  Generally, it is believed that,
in carbon rich  planetary nebulae,  dust  is composed  of carbon-based
grains, e.g. graphite, anthracite,  amorphous carbon and/or policyclic
aromatic hydrocarbons, while  in  oxygen-rich planetary nebulae  it is
composed  of   oxygen-based   grains,     e.g. different     forms  of
silicates. However, observations  with the  Infrared Space Observatory
showed that some  nebulae around Wolf-Rayet   central stars have  both
features (carbon-  and oxygen-based)  in    their spectra (Waters   et
al.~\cite{Waters}).

The optical properties of the various materials are different and this
affects the  infrared emission from the dust.  The number of planetary
nebulae in which the carbon abundance has been estimated is only about
50 (Rola \& Stasi\'{n}ska~\cite{Rola}),  and even in these objects, it
is often very uncertain.  Therefore,  in our statistical approach,  we
must take into account the fact that the grain composition varies from
object to object  in an  unknown  manner. We have therefore   computed
models  with  different  grain   compositions:  graphite   (Draine  \&
Laor~\cite{DraineLaor}),     amorphous     carbon   (Rouleau        \&
Martin~\cite{Rouleau})    and     circumstellar  silicates  (David  \&
Pegourie~\cite{David}).

In view of the large  parameter space to consider,   we have chosen  a
reference  model   with    $M_{*}$\,=\,0.60\,M$_{\odot}$,  $M_{\mathrm
{neb}}$\,=\,0.2\,M$_{\odot}$, $v_{\mathrm {exp}}$\,=\,20\,km\,s$^{-1}$
and dust grains  composed of graphite, with  $m_{\mathrm d}/m_{\mathrm
g}$\,=\,7.5\,10$^{-4}$ (which corresponds  to  a dust-to-hydrogen mass
ratio  $m_{\mathrm  d}/m_{\mathrm  H}$\,=\,10$^{-3}$). In  the figures
presented below, we discuss differences with respect to this reference
model by changing   only one defining  parameter. In  the  panels with
label  a,  $M_{*}$     takes   the  values   0.58,  0.60,    0.62  and
0.64\,M$_{\odot}$; in  panels b, $M_{\mathrm {neb}}$\,=\,0.1,  0.2 and
0.4\,M$_{\odot}$;  in panels c,   $v_{\mathrm  {exp}}$\,=\,10, 20  and
40\,km\,s$^{-1}$;    in      panels   d,  $m_{\mathrm    d}/m_{\mathrm
H}$\,=\,10$^{-4}$,   10$^{-3}$, 10$^{-2}$  and 10$^{-1}$; finally,  in
panels e we   explore various dust compositions:  graphite,  amorphous
carbon  and circumstellar silicates. It is  then relatively easy, from
these figures,  to imagine the  behavior of a  model  of any arbitrary
combination of defining parameters.

All the sequences of models  have been computed   with a time step  of
500\,yr, until an age  of 10\,000\,yr was reached,  and each symbol in
the figures below  represents one model.  Therefore zones where points
accumulate   in  a  theoretical  diagram  correspond   to  zones where
observational points should accumulate too (if observational selection
does not discriminate against this  particular  zone).  It would  have
been more satisfactory  to  build  the models  until a  given  surface
brightness  was reached. This,    however, would result  in very  long
sequences for some  cases, since it is easily  shown that  the time to
reach a given surface  brightness  goes like $M_{\mathrm  {neb}}^{2/5}
v_{\mathrm {exp}}^{-1}$  when  the nebulae have reached  the optically
thin regime.


\section{The contamination of IRAS broad band fluxes by line emission 
in the model nebulae}

As discussed in Pottasch  et al.~(\cite{Pottasch84a},b) and Lenzuni et
al.~(\cite{Lenzuni}), the broad band  fluxes measured by IRAS are  not
only due  to dust emission.  There  is a contribution from atomic line
emission,  as seen directly  in  the  low resolution  IRAS  spectra of
planetary        nebulae     (see        e.g.        Pottasch       et
al.~\cite{Pottasch84b}). Because this  effect was especially important
in the 12$\mu$m band, Pottasch et al.~(\cite{Pottasch84a}) and Lenzuni
et al.~(\cite{Lenzuni}) were careful not to use  fluxes from this band
to derive  the  dust properties (or,  rather,  they used them  only as
upper    limits).  Some  later       studies,  however  (Zhang      \&
Kwok~\cite{Zhang},  Tajitsu  \&  Tamura~\cite{Tajitsu})  seem  to have
ignored this problem.

Volk~(\cite{Volk})  and Vil'koviskii \&   Efimov~(\cite{Vil'koviskii})
included the line  emission  in their model  computations  of the IRAS
fluxes.  Volk~(\cite{Volk}),   for  example,   showed  that   the IRAS
color-color diagram was  strongly  affected by  atomic  line emission,
principally in the 12\,$\mu$m  band. However, he  explored only a very
restricted parameter space, and it is interesting to see how the lines
affect the  IRAS fluxes    in  general  during the  planetary   nebula
evolution.

\begin{figure*} 
\resizebox{\hsize}{!}{\includegraphics{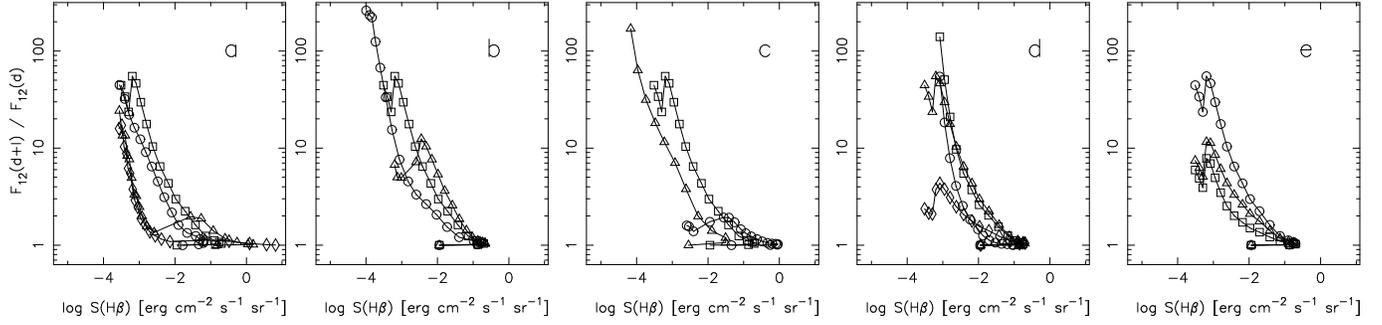}}
\caption{ 
The ratio $F_{\mathrm {12}}$(d$+$l)/$F_{\mathrm {12}}$(d) of
total  emission  and   pure  dust  emission in   the  12\,$\mu$m  band
vs. $S$(H$\beta$) for sequences of models computed at intervals of 500
yr  until 10\,000\,yr.  Each panel correspond  to the variation of one
parameter      with    respect    to      the   reference     model  (
$M_{*}$\,=\,0.60\,M$_{\odot}$,                             $M_{\mathrm
{neb}}$\,=\,0.2\,M$_{\odot}$, $v_{\mathrm {exp}}$\,=\,20\,km\,s$^{-1}$
and   graphite      grains        with   $m_{\mathrm     d}/m_{\mathrm
g}$\,=\,7.5\,10$^{-4}$).       {\bf  a:}     stellar  mass    $M_{*}$:
0.58\,M$_{\odot}$       (circles),    0.60\,M$_{\odot}$     (squares),
0.62\,M$_{\odot}$ (triangles), 0.64\,M$_{\odot}$ (diamonds).  {\bf b:}
nebular    mass $M_{\mathrm     {neb}}$:  0.1\,M$_{\odot}$  (circles),
0.2\,M$_{\odot}$ (squares),  0.4\,M$_{\odot}$   (triangles).  {\bf c:}
expansion velocity  $v_{\mathrm  {exp}}$: 10\,km\,s$^{-1}$  (circles),
20\,km\,s$^{-1}$ (squares),  40\,km\,s$^{-1}$ (triangles).    {\bf d:}
dust content $m_{\mathrm  d}/m_{\mathrm g}$: 7.5\,10$^{-2}$ (circles),
7.5\,10$^{-3}$  (squares), 7.5\,10$^{-4}$ (triangles),  7.5\,10$^{-5}$
(diamonds).   {\bf e:}   dust material:  graphite (circles), amorphous
carbon (squares), circumstellar silicates (triangles).  
}
\label{Fig1}
\end{figure*}

Figure~\ref{Fig1}a--e        shows       the      ratio    $F_{\mathrm
{12}}$(d$+$l)/$F_{\mathrm {12}}$(d) of the total
\begin{figure*} 
\resizebox{\hsize}{!}{\includegraphics{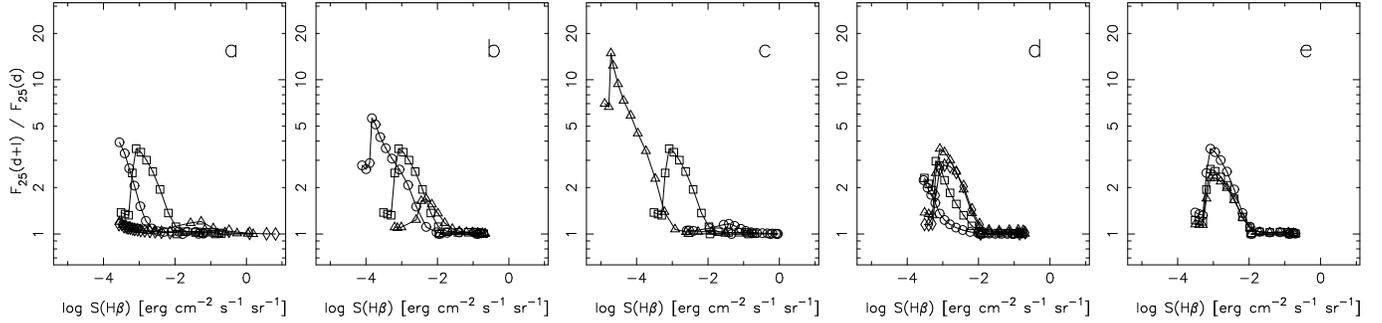}}
\caption{
Same as Fig.~\ref{Fig1} but for the ratio 
$F_{\mathrm {25}}$(d$+$l)/$F_{\mathrm {25}}$(d)
}
\label{Fig2}
\end{figure*}
\begin{figure*} 
\resizebox{\hsize}{!}{\includegraphics{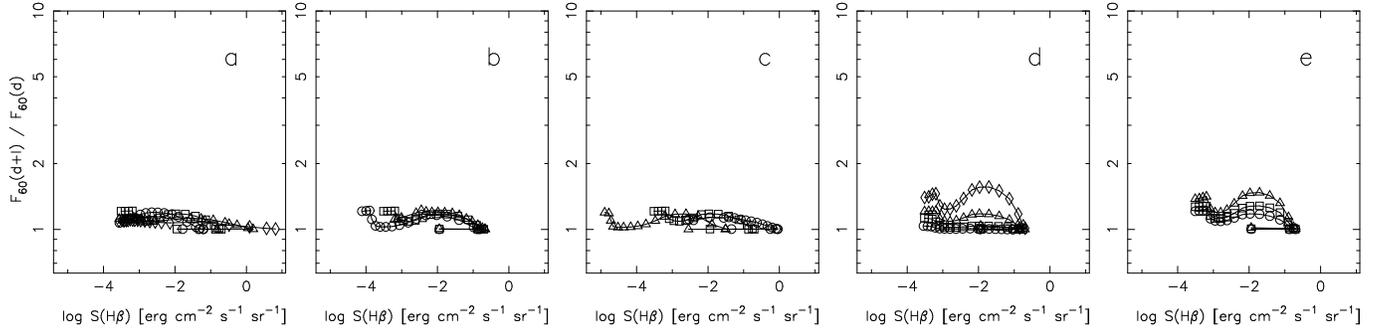}}
\caption{
Same as Fig.~\ref{Fig1} but for the ratio 
$F_{\mathrm {60}}$(d$+$l)/$F_{\mathrm {60}}$(d)
}
\label{Fig3}
\end{figure*}
\begin{figure*} 
\resizebox{\hsize}{!}{\includegraphics{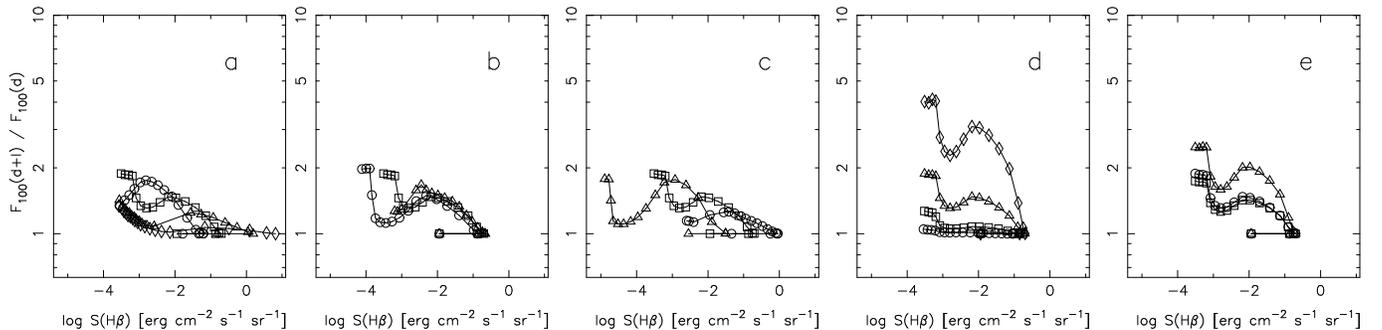}}
\caption{
Same as Fig.~\ref{Fig1} but for the ratio 
$F_{\mathrm {100}}$(d$+$l)/$F_{\mathrm {100}}$(d)
}
\label{Fig4}
\end{figure*}
emission in the  12\,$\mu$m band ({d$+$l} stands  for dust  $+$ lines)
and the emission due to dust only  (d), as a function of $S$(H$\beta$)
for   our   model       planetary   nebulae.   Figures~\ref{Fig2}a--e,
\ref{Fig3}a--e       and     \ref{Fig4}a--e    are   equivalent     to
figure~\ref{Fig1}a--e  for $F_{\mathrm {25}}$,  $F_{\mathrm {60}}$ and
$F_{\mathrm {100}}$, respectively.

We see   that,  as the nebula  ages,   the line   contribution  to the
$F_{\mathrm {12}}$ emission increases. This  is because, as the nebula
expands,   the  grains become cooler   and   gradually emit at  longer
wavelengths. With the  gas chemical composition  chosen in our models,
the lines that contribute   most to the  12\,$\mu$m band  are [Ne\,II]
12.8\,$\mu$m, then [S\,IV] 10.5\,$\mu$m, [Ar\,IV] 9.0\,$\mu$m, [Ar\,V]
7.9\,$\mu$m and  [Ne\,V]  14.3\,$\mu$m.  The latter  two  lines become
strong only  if the central star  is still luminous  when it reaches a
temperature above 150\,000\,K, which does not occurs for central stars
with $M_{*}$\,$\ge$\,0.62\,M$_{\odot}$.    Note        that  [Ne\,III]
15.5\,$\mu$m,  mentioned by  Pottasch et al.~(\cite{Pottasch84a}) does
not contribute to   the 12\,$\mu$m band. In  the   most extreme cases,
lines  may completely dominate  the flux in  the 12\,$\mu$m band. Even
for relatively young   objects,  the line contribution  may  still  be
important, depending  on the  characteristics  of the object  and  its
chemical composition.  Therefore,  unless a   direct  estimate of  the
infrared line strengths is available, one  should not use the observed
flux   in  the 12\,$\mu$m  band (except  as  an upper  limit) to infer
average  dust properties.  A further problem  with the 12\,$\mu$m band
is that it is contaminated by the emission from  the very small grains
that experience transient heating  which could be especially important
in the case of emission by policyclic aromatic hydrocarbons.

It is  interesting  to  note, in Fig.\,\ref{Fig1}d,  that  $F_{\mathrm
{12}}$(d$+$l)/$F_{\mathrm {12}}$(d) is the smallest in the sequence of
models with the lowest $m_{\mathrm d}/m_{\mathrm g}$. Intuitively, one
might  expect this ratio  to be larger  when the  amount  of dust in a
nebula is smaller. What  happens, in  such a  case, is that  the total
available energy to  heat the grains is shared  among a smaller number
of particles, making  each one hotter and the  result is a larger dust
emission at 12$\mu$m.

In  comparison with    the  $F_{\mathrm  {12}}$   emission,  the  line
contribution to the $F_{\mathrm {25}}$  emission is much smaller,  but
it can still dominate the dust emission  by up to  a factor about 3 in
the later stages of evolution,   as the  lines contributing most   are
[O\,IV] 25.9\,$\mu$m and [Ne\,V] 24.3\,$\mu$m.

The band with the smallest contribution  of atomic lines in our models
is the  60\,$\mu$m one ($<$\,50\,\% in  all cases), with  the lines of
[O\,III]    51.8\,$\mu$m    and [N\,III]    57.3\,$\mu$m  contributing
mostly. This band is much wider than the  12 and 25\,$\mu$m bands, and
much of the dust emission in planetary nebulae goes through this band.

\begin{figure} 
\resizebox{4.00cm}{!}{\includegraphics{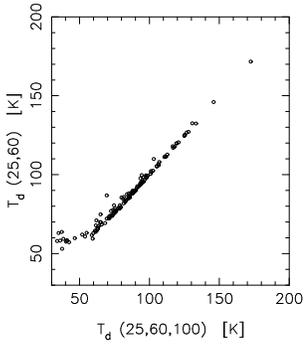}}
\caption{    
The  color   temperature  derived  from  the  $F_{\mathrm
{25}}/F_{\mathrm {60}}$ ratio as a function of the temperature derived
by a mean square fitting of $\lambda^{-1}\,B_{\lambda}(T_{\mathrm d})$
to the observed emission  in the three  bands 25, 60  and 100\,$\mu$m,
giving equal weight to all  three fluxes, for  our sample of planetary
nebulae.  
}
\label{Fig5}
\end{figure}

\begin{figure} 
\resizebox{4.00cm}{!}{\includegraphics{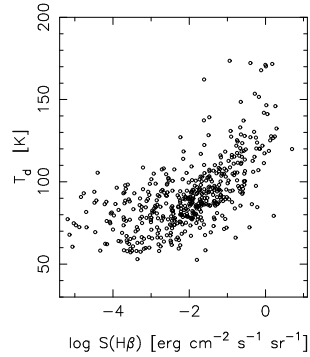}}
\caption{
The color temperature derived from the $F_{\mathrm {25}}/F_{\mathrm {60}}$ 
ratio as a function of $S$(H$\beta$) for our sample of planetary nebulae.
}
\label{Fig6}
\end{figure}

\begin{figure*} 
\resizebox{\hsize}{!}{\includegraphics{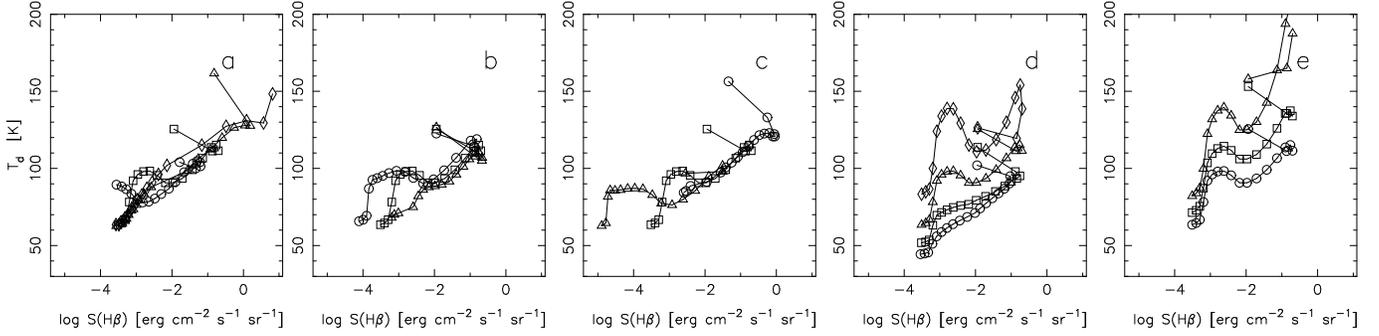}}
\caption{
The color temperature derived from the $F_{\mathrm {25}}$(d$+$l)/$F_{\mathrm 
{60}}$(d$+$l) ratios in the models as a function of $S$(H$\beta$). 
Same conventions as Fig.~\ref{Fig1}.
}
\label{Fig7}
\end{figure*}

The $F_{\mathrm  {100}}$ emission is   not too much affected  by  line
emission ([O\,III] 88.4\,$\mu$m  and  [N\,II] 121.9\,$\mu$m being  the
contributors) except at  a very low  dust-to-gas mass ratio  (of about
10$^{-4}$,  see  Fig.\,\ref{Fig4}d).  However,   it has  been reported
(IRAS Explanatory Supplement~\cite{IRASES})  that the 100\,$\mu$m band
may be contaminated  by cirrus emission  so that its use for planetary
nebulae may be  misleading unless a careful analysis  of  each case is
made individually.


\section{The dust temperature}

Characteristic dust temperatures are generally obtained by fitting the
observed spectrum   with either  $B_{\lambda}(T_{\mathrm  d}$),  where
$T_{\mathrm  d}$  is   the dust   temperature,  or   a function  which
approximates        the                      dust          emissivity:
$\lambda^{-1}\,B_{\lambda}(T_{\mathrm            d}$)               or
$\lambda^{-2}\,B_{\lambda}(T_{\mathrm  d}$)  (e.g.    Pottasch      et
al.~\cite{Pottasch84a},       Iyengar~\cite{Iyengar},     Zhang     \&
Kwok~\cite{Zhang}).      In    the   present      study,     we    use
$\lambda^{-1}\,B_{\lambda}(T_{\mathrm  d}$)  because  such  a function
better represents the   wavelength dependence of grain  emissivity for
the grains we have chosen.

That such  a quantity  as  a characteristic  dust temperature   can be
defined in a nebula may be surprising. Indeed, there is a distribution
of  the    temperatures  of  the  dust   grain   particles  inside the
nebulae.  Larger  grains have  smaller  temperatures,  since  they are
heated proportionally  to their surface  and radiate proportionally to
their volume. Also,   regions closer to the  central  star have higher
dust temperatures   because   the stellar   radiation   field is  less
diluted. Finally, grains located in the ionized zone of the nebula are
heated by  a harder radiation field than  those located in the neutral
zone where only the stellar photons at energies below 13.6\,eV and the
Ly$\alpha$ photons operate.  Nevertheless, our models show that, after
integrating   over  all the  grain  particles  in  a given nebula, the
resulting  infrared   spectrum    is   reasonably  indicative    of  a
characteristic dust temperature.  The  main parameter that affects the
dust temperature during the evolution of a model nebula is the average
distance of the dust particles from the  central star, which varies by
a factor 20 between 500 and 10\,000~yr.

To determine  the characteristic dust temperature  in the simplest way
one  can use  the ratio  of  the infrared  fluxes in   two bands only,
assuming that dust radiates like a single blackbody with an emissivity
proportional    to  $\lambda^{-1}$.   This    is  a  so-called   color
temperature.  We have tested on our  models that the color temperature
derived  from the $F_{25}$(d)/$F_{60}$(d)  ratio is very  close to the
temperature  derived   by      a    mean     square   fitting       of
$\lambda^{-1}\,B_{\lambda}(T_{\mathrm  d})$  to the pure dust emission
in the 25, 60 and 100\,$\mu$m bands,  giving equal weight to all three
fluxes   (which  confirms that the concept    of a characteristic dust
temperature is indeed a  valid one). Similarly, these temperatures are
also in  agreement in  the observed   planetary nebulae,  as  shown in
Fig.\,\ref{Fig5}, where the contribution of lines (and cirrus emission
in the  100\,$\mu$m band) adds some  slight dispersion. Using  a color
dust temperature based on 25 and 60$\,\mu$m fluxes allows us to derive
characteristic temperatures in a consistent way  for a large sample of
planetary  nebulae.  From now on,  $T_{\mathrm  d}$ will refer to  the
dust  temperature determined in such  a way.  In principle, one should
integrate  $\lambda^{-1}\,B_{\lambda}(T_{\mathrm  d})$  over the  IRAS
bands but we   have  checked that  doing  this results  in  increasing
$T_{\mathrm d}$  by less than  10\%. Column  6 of  Table 1  lists  the
values of $T_{\mathrm d}$ for all the planetary nebulae of our sample.

Figure~\ref{Fig6} shows the values of $T_{\mathrm d}$ as a function of
the nebular average surface brightness,  which has been computed using
the reddening  corrected H$\beta$  fluxes.  There is  a clear trend of
decreasing  $T_{\mathrm d}$  with   decreasing $S$(H$\beta$). This  is
reminiscent of the trend found by Pottasch et al.~(\cite{Pottasch84a})
and Lenzuni et al.~(\cite{Lenzuni}) of the dust temperature decreasing
with increasing nebular radius,  but the diagram  has the advantage of
being distance independent.  The dust temperatures in Fig.\,\ref{Fig6}
range   between  180\,K and  50\,K. A    few objects,  namely M\,2-23,
M\,1-48,  He\,2-34,  PC\,11,    He\,2-171, M\,2-6  and  H\,1-36,  have
$T_{\mathrm d}$ higher than 200\,K and do not appear in the plot.

Figure~\ref{Fig7}a--e shows   the  same as  Fig.\,\ref{Fig6}   for our
sequences of model planetary  nebulae, the  values of $T_{\mathrm  d}$
being  derived from the  emission in the 25  and 60$\,\mu$m bands {\em
including the lines}.  Roughly, the  models show a trend of decreasing
$T_{\mathrm d}$ with decreasing  surface brightness. This is explained
by a decreasing efficiency of the heating, mainly due to a dilution of
the   stellar radiation field  inside the  nebula as   it expands. The
models  reproduce the  observed diagram  remarkably  well, given their
simplicity.   Taking  all  the models   together,   one reproduces the
observed  range  in $T_{\mathrm d}$,  and  the observed variation with
$S$(H$\beta$). The   increase in   $T_{\mathrm    d}$ at  the   lowest
$S$(H$\beta$) is due  to an increasing contribution   of the lines  to
$F_{25}$, as seen in Fig.\,\ref{Fig2}.


\section{The dust-to-gas mass ratio}

The  formula used  by  Pottasch et  al.~(\cite{Pottasch84a}) to derive
dust-to-gas mass  ratios in planetary  nebulae  depends on the assumed
distance. We prefer using a formula that is distance independent.

The infrared luminosity at the wavelength $\lambda$ is:

\begin{eqnarray}
4\,\pi\,d^2\,F_{\lambda}\,=
\nonumber \\
\int_{a_{\mathrm {min}}}^{a_{\mathrm {max}}}\int_{R_{\mathrm {in}}}^{R_{\mathrm {out}}} 
K_{\lambda}^{\mathrm {abs}}(a) \rho_{\mathrm d}(r,a) 
B_{\lambda}(T_{\mathrm d}(r,a)) (4\,\pi\,r)^2 {\mathrm d}r\,{\mathrm d}a
\label{eq1}
\end{eqnarray}
where: $F_{\lambda}$ is  the measured  flux  from  the source at   the
distance $d$; $a_{\mathrm {min}}$  and $a_{\mathrm {max}}$ are minimum
and maximum dust grain sizes;  $K_{\lambda}^{\mathrm {abs}}(a)$ is the
mass absorption coefficient (which,   is almost independent of  $a$ at
infrared   wavelengths);    $\rho_{\mathrm  d}(r,a)$   is   the   dust
density. So, if $B_{\lambda}(T_{\mathrm d})$ can be considered uniform
in  the nebula and  independent of  $a$, which, as   we checked in our
models, is not a bad approximation, then the dust mass is simply given
by
\begin{equation}
m_{\mathrm d}\,=\,\frac{F_{\lambda} d^2}{K_{\lambda}^{\mathrm {abs}} 
B_{\lambda}(T_{\mathrm d})}
\label{eq2}
\end{equation}

The total luminosity in H$\beta$ is approximately given by: 
\begin{equation}
4\,\pi\,d^2\,F_{{\mathrm H}\beta}\,=\,\int_{R_{\mathrm 
{in}}}^{R_{\mathrm {out}}} 4.1\,10^{-22} T_{\mathrm e}^{-0.88}
n_{\mathrm e} n_{\mathrm H^+} 4 \pi r^2 {\mathrm d}r
\label{eq3}
\end{equation}
where the emissivity in H$\beta$ is given by Eq.\,IV--25 of 
Pottasch~(\cite{Pottasch84}); $T_{\mathrm e}$ and $n_{\mathrm e}$ are 
the electron temperature and density, respectively; $n_{\mathrm H^+}$ 
is the density of ionized hydrogen. If the gas is fully ionized and 
the  density is constant the total gas  mass is given by   
\begin{equation}
m_{\mathrm g}\,=\,\frac{1.4 m_{\mathrm H} 4 \pi d^2 F_{{\mathrm H}\beta}}
{4.1\,10^{-22} T_{\mathrm e}^{-0.88} n_{\mathrm e}}
\label{eq4}
\end{equation}
where $m_{\mathrm H}$ is the mass of the hydrogen atom and the factor 1.4
comes from the classical assumption of a He/H ratio of 0.1. For simplicity, 
we will assume for the remaining of the paper that $T_{\mathrm e}=10^{4}$K.

Therefore, using Eqs.\,(\ref{eq2}) and (\ref{eq4}) the dust-to-gas mass 
ratio is given by
\begin{equation}
m_{\mathrm d}/m_{\mathrm g}\,=\,\frac{F_{\lambda}}{F_{{\mathrm H}\beta}} 
\frac{1}{K_{\lambda}^{\mathrm {abs}}B_{\lambda}(T_{\mathrm d})} 
\frac{4.1\,10^{-22} T_{\mathrm e}^{-0.88} n_{\mathrm e}}{1.4 
m_{\mathrm H} 4 \pi}
\label{eq5}
\end{equation}
In principle, if  the colour temperature is  used, one can use equally
well the  fluxes in the 25\,$\mu$m  or the 60\,$\mu$m band to estimate
$m_{\mathrm   d}/m_{\mathrm g}$.   However,   since atomic  lines  may
contribute   to the observed IRAS  fluxes,  we prefer  to show results
based on the 60\,$\mu$m fluxes, since  this band is less contaminated,
as seen above (but the overall appearance of the diagrams based on the
25\,$\mu$m  fluxes is  very similar  to   the ones presented  in  this
paper).

\begin{figure*} 
\resizebox{\hsize}{!}{\includegraphics{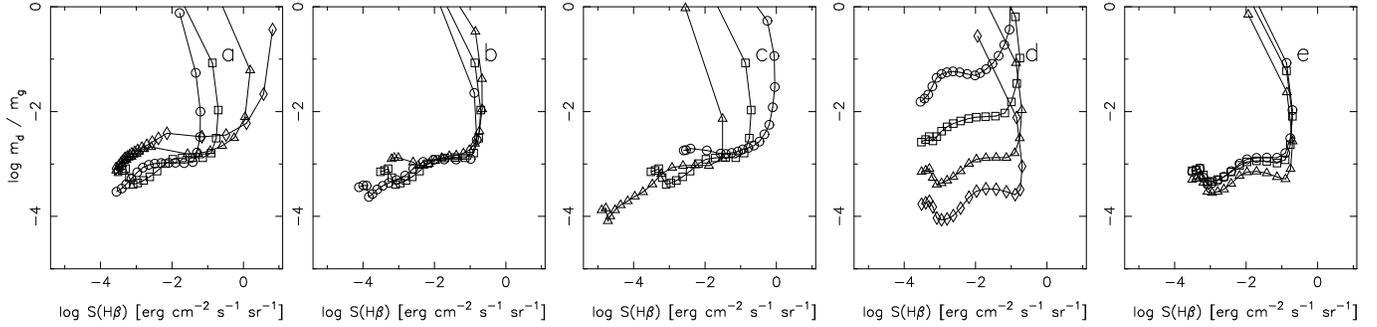}}
\caption{
The values of  $m_{\mathrm d}/m_{\mathrm g}$ derived from Eq.\,(\ref{eq5})
for the models as a function of $S$(H$\beta$). Same conventions as 
Fig.~\ref{Fig1}.
}
\label{Fig8}
\end{figure*}

To check the method before applying it to observed nebulae, we use it on
our models, where we know the input dust properties and dust-to-gas mass
ratio. Figure~\ref{Fig8}a--e shows the values of 
$m_{\mathrm d}/m_{\mathrm g}$ 
as derived   from  Eq.\,(\ref{eq5}),  as  a  function  of  $S({\mathrm
H}\beta$) for  our different series of  models. In  each case, we take
the value  of $K_{60}^{\mathrm {abs}}$  that corresponds {\em exactly}
to  the one used in the  model: 145.32  cm$^2$\,g$^{-1}$ for graphite,
74.38 for amorphous carbon and 53.45 for circumstellar silicates.

We find that, for many of the models, the formula returns an estimated
dust-to-gas mass ratio within   $\pm$\,0.3\,dex of the input one.   Of
course, the models    that are  optically    thick return  $m_{\mathrm
d}/m_{\mathrm  g}$ values  larger than   the   input ones, since   the
H$\beta$ flux  corresponds to the ionized region  only.  Most of those
optically  thick  models correspond  to  the  vertical  tracks seen in
Fig.  8 a--e, at the beginning  of the planetary nebulae evolution. As
seen  in  this  figure, the  total   number of model   points on these
vertical tracks is small compared to the number of points in the later
stages and they are located on the high surface brightness side of the
diagram.   Indeed, it  can  be   shown that,  whithin the  theoretical
framework        adopted      here,   planetary    nebulae        with
$M_{*}\,\leq$\,0.62\,M$_{\odot}$  are  in an ionization  bounded stage
only  during  a short      fraction  of their   observable   lifetime,
irrespective of  the   nebular  mass and  expansion   velocity.    For
planetary nebule whose central stars have  higher masses, however, the
proportion of  time spent in the  ionization bounded  regime is larger
(see  the figures in Stasi{\'n}ska  et al.~\cite{Stasinska98}), due to
the fact that the stellar luminosity drop occurs early in the lifetime
of the planetary nebula; the model with $M_*$\,=\,0.64\,M$_{\odot}$ in
Fig. 8a, corresponds to such a case.  There are however evidences that
the total number of planetary nebulae with  central stars more massive
than  0.62\,M$_{\odot}$ is not   large, 20\% at   the  very most  (see
Stasi{\'n}ska   et  al.~\cite{Stasinska97}).     We   note   also,  in
Fig.\,\ref{Fig8}, and especially in panels a--c, that, for some of the
very   low surface-brightness  models,    the use  of Eq.\,(\ref{eq5})
underestimates the  dust-to-gas mass ratio.  These cases correspond to
phases where the  atomic line contribution  to the 25\,$\mu$m  flux is
important,    as seen in    Fig.\,\ref{Fig2}a--e,  so that  the colour
temperature of      the  dust is   overestimated    which   induces an
underestimation of  the  dust-to-gas  mass ratio.   In  our series  of
models, this effect is  really strong only in  a few cases, mostly  at
the low surface brightness end.  However, the effect will be larger at
higher metallicities than considered in the models presented here.

It  is interesting  to compare the  range  in the  derived $m_{\mathrm
d}/m_{\mathrm g}$   values with  the range  of   the input  values  of
$m_{\mathrm   d}/m_{\mathrm   g}$    in   our   series   of    models.
Fig.\,\ref{Fig8}d  shows   that, if  one  excepts  the optically thick
cases,  the method tends  to  narrow the $m_{\mathrm d}/m_{\mathrm g}$
range with respect to the input one at  a given $S({\mathrm H}\beta)$.
This results from contamination of the  IRAS bands by atomic lines, as
can be  understood  by considering Figs.\,\ref{Fig2}   and \ref{Fig3}.
With  larger abundances of the metals  than in our  models, the effect
would be  even  more pronounced.  This  means that  the method itself,
when  applied  to  real nebulae,   does   not artificially widen   the
distribution of dust-to-gas mass ratios   with respect to  $S({\mathrm
H}\beta)$.

The sequence of    models   computed with  circumstellar     silicates
systematically returns a    smaller dust-to-gas mass    ratio than the
series  computed    with    graphite   or   amorphous   carbon  grains
(Fig.\,\ref{Fig8}e).  This   is  because  in    our  determination  of
$T_{\mathrm   d}$, we assumed that   the  dust emissivity varies  like
$\lambda^{-1}$, which  is indeed the  case  when comparing  the 25 and
60\,$\mu$m  wavelenghts for graphite  and  amorphous carbon, while for
circumstellar silicates  the    true   dependence is    rather    like
$\lambda^{-2}$. Therefore, we overestimate  the dust temperature, and,
consequently underestimate  the  dust mass.  This effect is  not  very
strong, about a factor 2 in $m_{\mathrm d}/m_{\mathrm g}$.  However if
we determine  the   dust-to-gas mass  ratio  of  the  model containing
silicates using the   value of $K_{60}^{\mathrm {abs}}$ for  graphite,
simulating the analysis of an object for which we do not know the true
value  of $K_{60}^{\mathrm  {abs}}$,  the estimated  dust-to-gas  mass
ratio goes down by an additional factor of about 2.

In  summary, combining all the  uncertainties, the method to determine
the dust-to-gas mass ratio  should give estimates  within, let us say,
$\pm$\,0.3\,dex in the majority of  planetary nebulae. As is known, it
overestimates $m_{\mathrm  d}/m_{\mathrm g}$ in  the case of planetary
nebulae that are   incompletely  ionized. Our  models  show that  such
cases, however, should represent a small fraction of observed nebulae,
unless one is dealing with samples at high  surface brightness. On the
other   hand, there is    probably  a tendency  to  underestimate  the
dust-to-gas mass   ratio for low surface    brightness objects, due to
contamination of the IRAS fluxes by  atomic lines, even when selecting
the best IRAS bands, as we have done here.

\begin{figure} 
\resizebox{4.00cm}{!}{\includegraphics{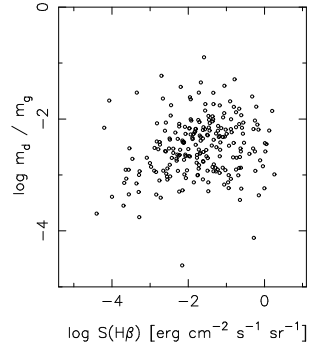}}
\caption{
The dust-to-gas ratio derived from Eq.\,(\ref{eq5}) 
as a function of $S$(H$\beta$) for our sample of planetary nebulae.
}
\label{Fig9}
\end{figure}

We  now turn to   the  observational sample.  Using  Eq.\,(\ref{eq5}),
where  $F_{{\mathrm H}\beta}$ now   represents the reddening corrected
nebular flux in  H$\beta$ and taking  for $K_{60}^{\mathrm {abs}}$ the
value  of 145.3 cm$^2$\,g$^{-1}$ corresponding  to graphite grains, we
have computed the values $m_{\mathrm  d}/m_{\mathrm g}$ for our sample
of planetary nebulae from the  observed fluxes in the 60\,$\mu$m band.
These $m_{\mathrm d}/m_{\mathrm g}$ are reported in  column 7 of Table
1.  Figure~\ref{Fig9}   shows  $m_{\mathrm  d}/m_{\mathrm   g}$  as  a
function of
$S({\mathrm H}\beta$).    We find  that   in  70\%  of the  cases  the
estimated  values  of   $m_{\mathrm   d}/m_{\mathrm  g}$ lie   between
10$^{-2}$  and 10$^{-3}$,   in  15\%  of  them they  are   larger than
10$^{-2}$ and in  15\%  of them they are  smaller  than 10$^{-3}$, the
extremes being close to  10$^{-1}$ and 10$^{-4}$.  Accounting for  the
fact that some planetary nebulae in the sample are ionization bounded,
and that atomic line contribution may  result in an underestimation of
$m_{\mathrm d}/m_{\mathrm g}$  in some low surface brightness objects,
the true range in $m_{\mathrm d}/m_{\mathrm g}$ is probably of about a
factor   10.  As a consequence,   the   dust-to-gas mass  ratio is   a
primordial  parameter for the interpretation of  far  infrared data of
planetary nebulae.   This was not  always taken into account in former
studies    of   statistical    nature  (Volk~\cite{Volk},   Zhang   \&
Kwok~\cite{Zhang}, Tajitsu \& Tamura~\cite{Tajitsu}).

Clearly, in Fig.  9   there  is  no  correlation  between  $m_{\mathrm
d}/m_{\mathrm g}$  and $S({\mathrm H}\beta$) (the  Pearson correlation
coefficient is 0.132 for 230 points, with an associated probability of
0.05).   Objects with very low or  very high $m_{\mathrm d}/m_{\mathrm
g}$ are found at any  $S({\mathrm H}\beta$).  The systematic errors in
the determination of $m_{\mathrm d}/m_{\mathrm g}$ that were discussed
above would tend to overestimate $m_{\mathrm d}/m_{\mathrm g}$ at high
$S({\mathrm  H}\beta$)   and   underestimate it   at  low  $S({\mathrm
H}\beta$).  But, as   we have argued above,  important  errors are not
likely  to affect a  large fraction of   the sample.  We conclude that
there is no evidence  for a decrease in  the dust-to-gas mass ratio as
the planetary nebulae evolve.


\section{The total infrared flux and the infrared excess}

Several   papers have studied  the   total infrared flux and  infrared
excess   in  planetary  nebulae. Pottasch et  al.~(\cite{Pottasch84a})
noted   that   younger planetary nebulae  have    higher IRE. Ratag et
al.~(\cite{Ratag90}) compared a sample of  about 100 planetary nebulae
in the galactic bulge and a sample of about 100 bright nearby nebulae,
and found   that  the latter  showed,  on   average,  lower IRE.  They
advocated that one possibility  could be for  example that, because of
selection effects, planetary nebulae  from the bulge sample would  be,
on  average, of higher surface  brightness and thus, younger.  Another
possibility suggested by Zijlstra et al.~(\cite{Zijlstra}) is that the
higher  IRE in the  bulge planetary nebulae could  be due  to a higher
dust content, related  to a higher metallicity  in the bulge. However,
Ratag et al.~(\cite{Ratag90})  rejected this option  since the average
metallicity of the bulge planetary  nebulae is not that much different
from the average metallicity of  nearby  disk planetary nebulae.  They
then  argued that the bulge  planetary nebulae,  being descendent of a
lower  mass stellar population,  would have lower  mass nuclei.  These
would spend  a longer  time  at lower effective temperature,  inducing
higher   IRE due to  the larger  contribution of  stellar non ionizing
photons to the  dust heating. Finally,  they  explored the possibility
that the high IRE in the galactic bulge planetary nebulae could be due
to additional dust heating  by the interstellar radiation field.  They
estimated that, for a planetary nebula located  in the galactic bulge,
an  energy  of 100  to   400  L$_{\odot}$ would   be  provided by  the
interstellar radiation field.

Our models are able to provide some insight into these questions, but let
us first discuss the empirical determination of the total infrared flux.

\begin{figure} 
\resizebox{4.00cm}{!}{\includegraphics{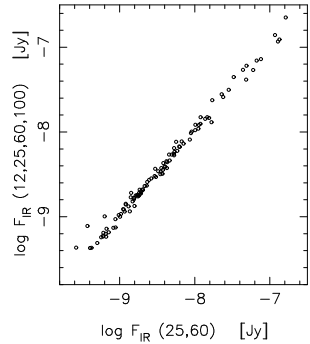}}
\caption{
The total infrared flux estimated from the IRAS fluxes at 12, 25, 60 
and 100\,$\mu$m as a function of the total infrared flux estimated from 
the IRAS fluxes at 25 and 60\,$\mu$m only
}
\label{Fig10}
\end{figure}

The   total  radiation that is  emitted  by  dust in a  nebula  can be
estimated from the IRAS fluxes basically by two methods. One is simply
to compute  the flux under the curve  defined by the observed infrared
fluxes, as done by Pottasch et al.~(\cite{Pottasch84a}).  The other is
to fit the observed energy  distribution by a function proportional to
${\lambda}^{-1}\,B_{\lambda}(T_{\mathrm d})$.  The simplest version of
this method is to  compute the blackbody  flux that corresponds to the
ratios  of the   observed fluxes  at 25  and   60\,$\mu$m.  Beside the
advantage of being  computationally simple, this  method allows one to
consider a large number of planetary nebulae, since it requires fluxes
in  only two bands, and it  is the least  affected  by atomic line and
cirrus emission. It is interesting to compare the results of these two
methods. This  is done  in Fig.\,\ref{Fig10}  where we plot  the total
IRAS
flux determined by the first method using the fluxes at 12, 25, 60 and
100\,$\mu$m as  a  function of the total   IRAS  flux derived  by  the
simplest version of the second  method. In  building this diagram,  we
considered only  those objects with $Q$\,=\,3  in the four IRAS bands,
so that  the  total number  of objects represented  is   only 107. The
observed correlation is very good over the whole range of fluxes. From
this,  we feel confident   that we  can  indeed  use  only the 25  and
60\,$\mu$m  fluxes to estimate the total  infrared flux.  The infrared
excess is then  determined from the  total infrared flux and the total
reddening corrected H$\beta$ flux using
\begin{equation}
{\mathrm {IRE}}\,=\,\frac{F_{\mathrm {IR}}}{23.3\,F_{{\mathrm H}\beta}}
\label{eq6}
\end{equation}
where 23.3 is the ratio of Ly$\alpha$ to H$\beta$ emission (Table\,VIII-2 
of Pottasch~\cite{Pottasch84}).

\begin{figure} 
\resizebox{4.00cm}{!}{\includegraphics{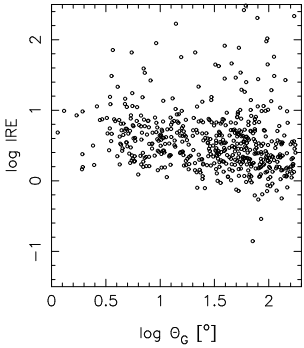}}
\caption{
The values of IRE derived from Eq.\,(\ref{eq6}) as a function of the 
angular distance to the Galactic center for our sample of planetary 
nebulae.
}
\label{Fig11}
\end{figure}
Figure~\ref{Fig11} shows   the IRE, determined   in such  a  way, as a
function of the angular distance to the Galactic center for our sample
of  planetary nebulae.  The sample  is much larger  than considered in
previous  studies since it contains  more  than 500 objects.  There is
indeed a  tendency for  the average infrared   excess to decrease with
increasing  angular  distance to  the  Galactic center.   However, the
variation  is  not  exactly  the one that  would  be   expected if the
interstellar radiation   field alone were  to produce  this effect, as
suggested by Ratag et al.~(\cite{Ratag90}) and Amnuel~(\cite{Amnuel}).
Indeed, the variation is  very smooth and  shows no marked peak in the
direction of the bulge. We  also note the  existence of objects (about
10\% of the whole  sample) with IRE $>$ 10,  and those are distributed
uniformly in Galactic longitude. This   indicates that these high  IRE
are related to an  intrinsic property of the nebulae  and not to their
location in the Galaxy.

\begin{figure} 
\resizebox{4.00cm}{!}{\includegraphics{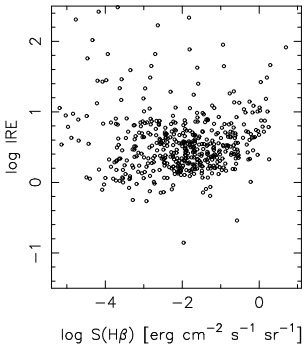}}
\caption{
The values of IRE derived from Eq.\,(\ref{eq6}) as a function of 
$S$(H$\beta$) for our sample of planetary nebulae.
}
\label{Fig12}
\end{figure}
Figure~\ref{Fig12}  shows the IRE as a  function  of $S$(H$\beta$). In
this figure, which  contains over 470  points (the angular diameter is
required in addition to the H$\beta$ flux), one can see that planetary
nebulae    of high or low  surface   brightness can  have equally high
IRE. There is  no  evidence that the IRE   is higher  for  the younger
nebulae.

\begin{figure*} 
\resizebox{\hsize}{!}{\includegraphics{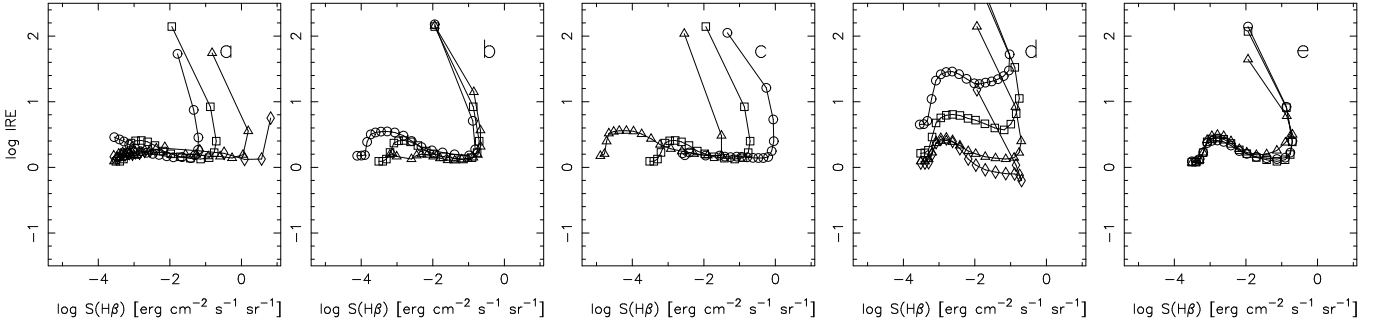}}
\caption{
The values of IRE derived from Eq.\,(\ref{eq6}) for the models as a 
function of $S$(H$\beta$). Same conventions as Fig.~\ref{Fig1}.
}
\label{Fig13}
\end{figure*}

\begin{figure*} 
\resizebox{\hsize}{!}{\includegraphics{MS8911.f14}}
\caption{
The values of IRE derived from Eq.\,(\ref{eq6}) for the models as a 
function of  $T_{\mathrm {eff}}$. Same conventions as Fig.~\ref{Fig1}.
}
\label{Fig14}
\end{figure*}
Figure~\ref{Fig13}a--e    represents         the  same   diagram    as
Figure~\ref{Fig12} for  our sequences of  models.  It shows  that, for
the earliest evolutionary stages, when  the central star does not emit
a large proportion of ionizing photons, the infrared excess can indeed
reach  very high   values,   as  was   argued in   several    previous
studies.  But, according to  our  models, such optically thick objects
should be rare. On the other hand, large  values of the IRE are easily
produced     when    $m_{\mathrm  d}/m_{\mathrm     g}$    is     high
(Fig.\,\ref{Fig13}d).  Since  we   have  shown that  large dust-to-gas
ratios occur for planetary nebulae  of all ages, the  same is true for
large values of the IRE.

It has  been suggested by Zijlstra  et al.~(\cite{Zijlstra})  that the
infrared  excess may be  used as a  measure of the  temperature of the
central stars,  $T_{\mathrm  {eff}}$, for stars  cooler  than 40\,000K
(for  larger values of   $T_{\mathrm {eff}}$, the proposed calibration
becomes  multivalued).  Figure~\ref{Fig14}a--e  shows  the  IRE   as a
function of $T_{\mathrm {eff}}$ in our  sequences of models. From this
figure, we see  that the method is  less reliable than it appears from
Fig.  B1 of Zijlstra et al.~(\cite{Zijlstra}).   Even  for the coolest
stars, for  a given $T_{\mathrm  {eff}}$, the IRE  depends somewhat on
the  mass  of the  central star,  on the  nebular  mass  and expansion
velocity and,  most importantly, on the  dust-to-gas  mass ratio. As a
result, the derivation of   $T_{\mathrm  {eff}}$ using  the   infrared
excess  is  not  accurate. It  will  be especially  misleading if  the
dust-to-gas  mass  ratio  in   a given  nebula    does not allow   the
simplifying assumptions made by  Ziljstra et al. ~(\cite{Zijlstra}) to
be valid.


\section{The grain size}

Lenzuni et al.~(\cite{Lenzuni}) estimated the typical size of the dust
grains in their sample from   the thermal balance equation, under  the
assumption that the absorption efficiency  $Q_{\mathrm {UV}}$ was 1 in
the UV and   that they could   estimate the  total  luminosity of  the
central   star. They  found  that  the   average size of  the  grains,
$\overline{a}$, was    decreasing   from about  3.\,10$^{-5}$   cm  to
3.\,10$^{-8}$  cm as    the  radius of  the    nebulae increased  from
3.\,10$^{16}$ to 1.\,10$^{18}$  cm. This was an  appealing conclusion,
which,  combined with  their finding that  the  total dust-to-gas mass
ratio  was decreasing with time, led  to some interesting speculations
about the destruction process of dust grains in planetary nebulae.

Actually, it is not quite true that $Q_{\mathrm {UV}}$ is constant, it
varies    strongly with wavelength    for   grains smaller than  about
10$^{-6}$ -- 5.\,10$^{-6}$ cm depending on the dust material.

\begin{figure*} 
\resizebox{\hsize}{!}{\includegraphics{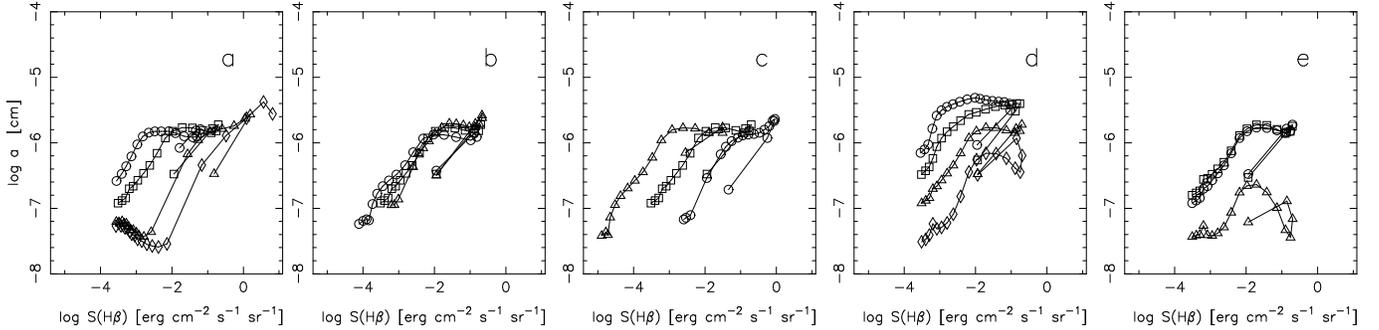}}
\caption{
The values of the average grain size $\overline{a}$ derived from 
Eq.\,(\ref{eq7} for the models as a function of $S$(H$\beta$). 
Same conventions as Fig.~\ref{Fig1}.
}
\label{Fig15}
\end{figure*}

The method of Lenzuni et  al.~(\cite{Lenzuni}) can easily be tested on
our models, which all have the same distribution of grain sizes during
the whole  nebular evolution.  The  formula they used can be rewritten
as:
\begin{equation}
\overline{a}\,=\, 
\frac{3\,\lambda^{-1}\,Q_{\mathrm {UV}}\,L_*}
{(4\,\pi\,r)^2\,4\,\rho_{\mathrm d}^{\mathrm {bulk}} 
K_{\lambda}^{\mathrm {abs}}
\int_0^{\infty} \lambda^{-1}
B_{\lambda}(T_{\mathrm d}) {\mathrm d}\lambda}
\label{eq7}
\end{equation}
where $L_*$ is the stellar luminosity and $\rho_{\mathrm d}^{\mathrm 
{bulk}}$ the density of the grain material. In the case of our models, 
we know the total stellar luminosity, so the formula can be applied 
easily. Fig.\,\ref{Fig15}a--e shows the resulting value 
of   $\overline{a}$      obtained     using   Eq.\,(\ref{eq7})      at
$\lambda\,=\,60\,\mu$m for our sequences  of models.  We see that  the
method to    derive  $\overline{a}$  is strongly  biased,    and gives
decreasing  values of $\overline{a}$   as $S$(H$\beta$) decreases. The
effect is quite large:  the estimated $\overline{a}$ decrease by about
2\,dex as  $S({\mathrm H}\beta$) decreases  by about 4\,dex  which, in
our models,  correspond roughly to an  increase of the radius by about
1\,dex. Therefore, the  bias in  the method  is of  the same order  of
magnitude as the effect found by Lenzuni et al.~(\cite{Lenzuni}).

We  have   not   questioned  here the     way   in which   Lenzuni  et
al.~(\cite{Lenzuni}) determined  $L_*$. They  used an analytical model
of  HII regions (Panagia~\cite{Panagia})   to   do this.  This   model
involves many assumptions that are not necessarily fulfilled and might
produce an additional bias in one way or another.

Since, in  our models, the value  of $\overline{a}$ is 3.53\,10$^{-6}$
cm,  we see   from Fig.\,\ref{Fig15} that   the method  of Lenzuni  et
al. rather tends to underestimate $\overline{a}$.  If we consider that
the highest  values  of $\overline{a}$ found   by  Lenzuni et al.  are
$\simeq$ 3.\,10$^{-5}$  cm, it might be that  indeed such large grains
are found in planetary nebulae.  Such a conclusion, however, might not
be robust, since we  have not tested how  it would vary if we  assumed
different size distributions or forms of the grains. Interestingly, we
note that, in the Red Rectangle nebula, Jura~(\cite{Jura}) has argued,
on other grounds, that grains have a large size.

The  main conclusion  from  testing Eq.\,(\ref{eq7}) on  our models is
that, clearly, one cannot derive the value of $\overline{a}$ in such a
way. One could, perhaps estimate this  parameter from a detailed model
fitting of individual, well observed, objects, but this is not certain
and has never  been attempted so far  to  our knowledge. For the  time
being, we consider  that there is  no evidence that the  typical grain
size of dust varies during the evolution of planetary nebulae.


\section{Discussion}

We  have    reconsidered the dust   content  of  planetary  nebulae by
performing a new statistical study of broad band  IRAS data.  The main
difference with  previous studies is that our   method of analysis was
tested on   a wide grid  of photoionization  models  of evolving dusty
planetary  nebulae. Furthermore, our  results   are based on  distance
independent diagrams.  We are  aware that our analysis, although  more
elaborated than previous studies of  statistical nature, still suffers
from uncertainties.  Detailed modelling  of individual objects should,
in   principle,  provide better  estimates  of  the   dust  content of
planetary nebulae.   Such  an approach is  however  not feasible  on a
large sample of objects.

Our main result  is that the  range  in dust-to-gas mass  ratios among
planetary nebulae is large, about  a factor 10,  but, in contrast with
previous authors, we find no evidence for a  change in the dust-to-gas
mass ratio in the course of planetary nebulae evolution. Neither do we
find evidence for a change in the average grain size.
 
To our knowledge, the problem of dust survival in planetary nebulae is
still  poorly understood from  a theoretical  point  of view. The main
processes    of  dust destruction have  been    described by Draine \&
Salpeter~(\cite{DraineSalpeter}       and  references  therein)     or
Draine~(\cite{Draine90}) but  the estimated efficiencies are uncertain
and need confirmation by observation.  After  the works of Pottasch et
al.~(\cite{Pottasch84a}) and Lenzuni  et al.~(\cite{Lenzuni}) the idea
that  dust is gradually destroyed  -- by the energetic photons emitted
by the  central star  or by other  mechanisms --  made  its  way among
astronomers  (e.g. Furton  \&  Witt~\cite{Furton}, Dinerstein  et
al.~\cite{Dinerstein},   Corradi    \& Schwarz~\cite{Corradi}).    Our
analysis  does not  support such a  view.   In addition, there  is now
observational evidence  that dust destruction  in planetary nebulae is
probably   not  as  strong  as    thought  before.  Namely,  Volk   et
al.~(\cite{Volk97}) have reported   that their attempt to  search  for
lines of Ca  in planetary nebulae   has failed, indicating that Ca  is
probably  locked      in   grains  (see   the    classical    work  of
Field~\cite{Field}   which   discusses the   problem of  the gas-phase
element  depletions) and   the UV radiation  does   not seem to be  so
efficient  for dust destruction. It has  been shown by the theoretical
calculations   of   Okorokov et   al.~(\cite{Okorokov}) and  Marten et
al.~(\cite{Marten})  that   radiation    pressure on   dust  particles
accelerates particles to  higher velocities and gradually expells them
from the nebula. It seems however, that the  conclusion of Okorokov et
al.~(\cite{Okorokov})  that  dust  is  removed from  planetary nebulae
almost entirely is too far going. They did not consider the process of
charging of the  dust particles which probably  freezes them  into the
ionized gas.  It is worth  of noting, however, that  the spread of the
points representing the dust-to-gas mass ratio is quite wide and shows
that planetary with different dust contents exist. The reason could be
at least twofold: planetary  nebulae  could be created  with different
dust-to-gas mass  ratios  on the AGB; or  the  time spent by  the star
during the pre-planetary phase of evolution, when  the process of dust
acceleration  and removing is  efficient,  was different (shorter  for
planetary nebulae with   presently higher dust-to-gas  mass ratio). If
the second  scenario applies, one can  speculate that, at least from a
statistical point of view,  planetary nebulae with smaller dust-to-gas
ratio might have less massive central stars.


\begin{acknowledgements}

We thank S. G\'{o}rny for having made available his compilation of 
electron densities in planetary nebulae and J. Mathis for his precious 
comments about the manuscript. Remarks by an anonymous referee were
helpful in clarifying some aspects of this work.
This work is in part supported by the grant 2.P03D.002.13 to R.S. from
the Polish State Committee for Scientific Research, by CNRS: ``jumelage
France--Pologne'', by the University Paris 7 and by the Observatoire de 
Paris-Meudon.

\end{acknowledgements}


{}

\end{document}